\documentclass{article}

\usepackage{spconf,amsmath,graphicx}
\usepackage{amsfonts, amsopn, amsthm}

\title{A Double Joint Bayesian Approach for J-Vector Based Text-dependent Speaker Verification}
\name{Ziqiang Shi, Mengjiao Wang, Liu Liu, Huibin Lin, Rujie Liu}
\address{Fujitsu Research and Development Center, Beijing, China}
\begin{document}
%\ninept
%
\maketitle
\begin{abstract}
J-vector has been proved to be very effective in text-dependent speaker verification with short-duration speech. However, the current state-of-the-art back-end classifiers, e.g. joint Bayesian model, cannot make full use of such deep features. In this paper, we generalize the standard joint Bayesian approach to model the multi-faceted information in the j-vector explicitly and jointly. In our generalization, the j-vector was modeled as a result derived by a generative Double Joint Bayesian (DoJoBa) model, which contains several kinds of latent variables.
With DoJoBa, we are able to explicitly build a model that can combine multiple heterogeneous information from the j-vectors.
In verification step, we calculated the likelihood to describe whether the two j-vectors having consistent labels or not.
On the public RSR2015 data corpus, the experimental results showed that our approach can achieve 0.02\% EER and 0.02\% EER for impostor wrong and impostor correct cases respectively.
\end{abstract}
\section{Introduction}
\label{sec:intro}

As opposed to text-independent speaker verification, where the speech content is unconstrained, text-dependent speaker verification systems are more favorable for security applications since they showed higher accuracy on short-duration sessions~\cite{larcher2014text,Heigold2016End}.

The previous methods regarding text-dependent speaker verification can be grouped into two categories. The first category is based on the traditional state-of-the-art GMM-UBM or i-vector approach, which may not work well in this case~\cite{kenny2014jfa,larcher2014text,dehak2011front}.
In the second category, deep models are ported to speaker verification: deep neural network (DNN) is used to estimate the frame posterior probabilities~\cite{lei2014novel}; DNN as a feature extractor for the utterance representation~\cite{Variani2014Deep}; Matejka et al.~\cite{Zeinali2016i} have shown that using bottle-neck DNN features (BN) concatenated to other acoustic features outperformed the DNN method for text-dependent speaker verification; multi-task learning jointly learns both speaker identity and text information~\cite{chen2015multi}.

This paper is based on two works: one is of Chen et al.~\cite{chen2015multi}, in which the j-vector was introduced as a kind of more compact representation for text dependent utterances; the other is of Chen et al.~\cite{chen2017efficient}, in which the state-of-the-art joint Bayesian analysis is proposed to model the two facial images jointly with an appropriate prior that considers intra- and extra-personal variations over the image pairs. However, the standard joint Bayesian model only considers one single label, but in practice the extracted features are always associated with several labels, for example when using multi-task learned networks as feature extractor to extract the j-vector~\cite{chen2015multi}.
Since j-vector potentially have different kinds of labels, the text latent variable is no longer only dependent on the current label, but rather depends on a separate text label. This means for j-vector there are two latent variables related to speaker and text have equal importance, and both variables are tied across all samples that sharing a certain label.

In order to better modeling j-vector, we propose a generalization of the standard joint Bayesian~\cite{chen2017efficient} called Double Joint\footnotemark[1] Bayesian (DoJoBa), which can explicitly and jointly model the multi-view information from samples, such as certain individual saying some text content. The relationship between DoJoBa and standard joint Bayesian is analogous to that between joint factor analysis and factor analysis. DoJoBa is also related to the work of Shi et al.~\cite{Shi2017Multi}, in which a joint PLDA is proposed for j-vector verification. One of the most important advantages of DoJoBa compared to joint PLDA, is that DoJoBa can learn the appropriate dimensionality (or the number of columns) of the low-rank speaker subspace and phrase subspaces without user tuning.

\footnotetext[1]{For the ``double joint'' term, the first ``joint'' is for modeling the multi-view information jointly, e.g. text and identity in j-vector, while the second ``joint'' is for joint distribution of two features, e.g. target and test j-vectors}

The remainder of this paper is organized as follows: Section 2 reviews the standard j-vector/joint Bayesian system. Section 3 describes the DoJoBa approach. The detailed experimental results and comparisons are presented in Section 4 and the whole work is summarized in Section 5.

\section{Baseline j-vector/joint Bayesian model}
\label{sec:baseline}

The standard j-vector~\cite{chen2015multi} and the joint Bayesian model~\cite{chen2017efficient} is used as the baseline in this work. This section gives a brief review of this baseline.

\subsection{J-vector extraction}

Chen et al.~\cite{chen2015multi} proposed a method to train a DNN to make classifications for both speaker and phrase by minimizing a total loss function consisting a sum of two cross-entropy losses - one related to the speaker label and the other to the text label. Once training is completed, the output layer is removed, and the rest of the neural network is used to extract speaker-phrase joint features. Each frame of an utterance is forward propagated through the network, and the output activations of all the frames are averaged to form an utterance-level feature called j-vector. The enrollment speaker models are formed by averaging the j-vectors corresponding to the enrollment recordings.

\subsection{The joint Bayesian model}

For the back-end, the state-of-the-art joint Bayesian model~\cite{chen2017efficient} is employed as a classifier for speaker verification. For simplicity of notation, joint Bayesian model with only single speaker label is used here as an example. Joint Bayesian models data generation using the following equation:
\begin{equation}
x_{ij} = \mu + z_i + \epsilon_{ij} \nonumber
\end{equation}
where $x_{ij}$ is certain j-vector, and $z_i$ and $\epsilon_{ij}$ are defined to be Gaussian with diagonal covariance $\Sigma_z$ and $\Sigma_\epsilon$ respectively.

The parameters $\theta=\{ \mu, \Sigma_z, \Sigma_\epsilon\}$ of this joint Bayesian model can be estimated using the Expectation Maximization (EM)~\cite{Dempster1977Maximum,chen2017efficient} algorithm.
With the learned joint Bayesian model, given a test $x_t$ and an enrolled model $x_s$, the likelihood ratio score is
\begin{equation*}
l(x_t,x_s) = \frac{P(x_t,x_s|\text{same-speaker})}{P(x_t,x_s|\text{different-speakers})}.
\end{equation*}

This standard joint Bayesian cannot properly deal with the j-vector that jointly belong to certain speaker and certain phrase at the same time. For j-vector, it is noted that we need to define the joint Bayesian latent variable $z_i$ as the joint variable considering both speaker and phrase information. This means the latent variable $z_i$ is dependent on both a speaker identity and a phrase label. In this work we try to separate the $z_i$ into two independent latent variables - one related to the speaker identity information and the other to the phrase.
This intuitive idea results in the following DoJoBa.

\section{Double joint Bayesian model}
\label{sec:dojoba}
%Contrary to the baseline PLDA model, the new one is defined by a
In this section, we propose an effective model to describe the j-vector as resulting from a generative model which incorporates both intra-speaker/phrase and inter-speaker/phrase variation.

\subsection{Generative model}

We assume that the training data is obtained from $I$ speakers saying $J$ phrases each with $H_{ij}$ sessions. We denote the j-vector of the $k$'th session of the $i$'th speaker saying $j$'th phrase by $x_{ijk}$. We model the text dependent feature generation by the process:
\begin{equation}
\label{eq:dojoba}
x_{ijk} = \mu + u_i + v_{j} + \epsilon_{ijk}.
\end{equation}

The model comprises two parts: 1, the signal component $\mu + u_i + v_{j}$  which depends only on the speaker and phrase, rather than on the particular feature vector (i.e. there is no dependence on $k$); 2, the noise component $\epsilon_{ijk}$ which is different for every feature vector of the speaker/phrase and represents within-speaker/phrase noise. The term $\mu$ represents the overall mean of the training vectors. Remaining unexplained data variation is explained by the residual noise term $\epsilon_{ijk}$ which is defined to be Gaussian with diagonal covariance $\Sigma_\epsilon$.
The latent variables $u_i$ and $v_{j}$ are defined to be Gaussian with diagonal covariance $\Sigma_u$ and $\Sigma_v$ respectively, and are particularly important in real application, as these represents the identity of the speaker $i$ and the content of the text $j$ respectively.

Formally the model can be described in terms of conditional probabilities
\begin{eqnarray*}
p(x_{ijk}|u_i,v_j, \theta) &=& \mathcal{N}(x_{ijk}| \mu + u_i + v_{j}, \Sigma_\epsilon), \\
p(u_i) &=&  \mathcal{N}(u_i|0,\Sigma_u), \\
p(v_j) &=&  \mathcal{N}(v_j|0,\Sigma_v).
\end{eqnarray*}
where $\mathcal{N}(x|\mu,\Sigma)$ represents a Gaussian in $x$ with mean $\mu$ and covariance $\Sigma$.
Here it's worth to notice that the mathematical relationship between DoJoBa and joint Bayesian~\cite{chen2017efficient} is analogous (not exactly) to that between joint PLDA~\cite{Shi2017Multi} and PLDA~\cite{Jiang2012PLDA}. Compared to joint PLDA, DoJoBa allows the data to determine the appropriate dimensionality of the low-rank speaker and text subspaces for maximal discrimination, as opposed to requiring heuristic manual selections.

Let $X=\{x_{ijk}\in \mathbb{R}^D: i=1,...,I;j=1,...,J;k=1,...,H_{ij}\}$, $x_{ij}=\{x_{ijk}:k=1,...,H_{ij}\}$, and  $x_i=\{x_{ijk}:j=1,...,J;k=1,...,H_{ij}\}$.
In order to maximize the likelihood of data set $X$ with respect to parameters $\theta=\{\mu, \Sigma_u, \Sigma_v, \Sigma_\epsilon\}$, the classical EM algorithm~\cite{Dempster1977Maximum} is employed.

\subsection{EM formulation}
 The auxiliary function for EM is
\begin{align*}
 &Q(\theta|\theta_t) = \text{E}_{U,V|X,\theta_t}[\log p(X,U,V|\theta) ] \\
=&   \text{E}_{U,V|X,\theta_t}\left\{\sum_{i=1}^I\sum_{j=1}^{J}\sum_{k=1}^{H_{ij}}\log [p(x_{ijk}|u_{i},v_{j}, \theta)p(u_i,v_j)] \right\}  \nonumber
\end{align*}

By maximizing the auxiliary function, we obtain the following EM formulations.

\textbf{E} steps:we need to calculate the expectations $\text{E}_{U|X,\theta_t}[u_i]$, $\text{E}_{V|X,\theta_t}[v_j]$, $\text{E}_{U|X,\theta_t}[u_iu_i^T]$, $\text{E}_{V|X,\theta_t}[v_jv_j^T]$, and  $\text{E}_{U,V|X,\theta_t}[u_iv_j^T]$.
Indeed we have
\begin{align}
\label{eq:estep1}
&\text{E}_{U|X,\theta_t}[u_i]= \\ &\left(\Sigma_u^{-1}+\Sigma_\epsilon^{-1}\sum_{j=1}^{J}H_{ij}\right)^{-1}{\Sigma}_\epsilon^{-1}\sum_{j=1}^{J}\sum_{k=1}^{H_{ij}}(x_{ijk}-\mu-v_j). \nonumber
\end{align}
and
\begin{align}
\label{eq:estep2}
&\text{E}_{U|X,\theta_t}[u_iu_i^T]= \\ &\left(\Sigma_u^{-1}+\Sigma_\epsilon^{-1}\sum_{j=1}^{J}H_{ij}\right)^{-1}+\text{E}_{U|X,\theta_t}[u_i]\text{E}_{U|X,\theta_t}[u_i]^T. \nonumber
\end{align}
It is almost the similar equations for $\text{E}_{V|X,\theta_t}[v_j]$ and $\text{E}_{V|X,\theta_t}[v_jv_j^T]$.
For $\text{E}_{U,V|X,\theta_t}[u_iv_j^T]$, we have
\begin{align}
\label{eq:estep3}
&\text{E}_{U,V|X,\theta_t}\left\{\left[\begin{matrix}
    u_iu_i^T & u_iv_j^T \\ v_ju_i^T & v_jv_j^T
    \end{matrix}\right]\right\}= \\  & \left(\textbf{diag}[\Sigma_u^{-1},\Sigma_v^{-1}]+H_{ij}\mathbf{B}^T{\Sigma}_\epsilon^{-1}\mathbf{B}\right)^{-1}\nonumber\\+&\text{E}_{U,V|X,\theta_t}\left\{\left[\begin{matrix}
    u_i \\ v_j
    \end{matrix}\right]\right\}\text{E}_{U,V|X,\theta_t}\left\{\left[\begin{matrix}
    u_i \\ v_j
    \end{matrix}\right]\right\}^T\nonumber
\end{align}
where $\mathbf{B}=\left[\begin{matrix}
    \mathbf{I} & \mathbf{I}
    \end{matrix}\right]$ and
\begin{align*}
&\text{E}_{U,V|X,\theta_t}\left\{\left[\begin{matrix}
    u_i \\ v_j
    \end{matrix}\right]\right\}= \\ & \left(\textbf{diag}[\Sigma_u^{-1},\Sigma_v^{-1}]+H_{ij}\mathbf{B}^T{\Sigma}_\epsilon^{-1}\mathbf{B}\right)^{-1}\mathbf{B}^T{\Sigma}_\epsilon^{-1}\sum_{k=1}^{H_{ij}}(x_{ijk}-\mu).
\end{align*}

\textbf{M} steps: we update the values of the parameters $\theta=\{\mu, \Sigma_u, \Sigma_v, \Sigma_\epsilon\}$ and have
\begin{equation*}
\Sigma_u =  \frac{1}{\sum_{i=1}^I\sum_{j=1}^{J}\sum_{k=1}^{H_{ij}}1}\sum_{i=1}^I\sum_{j=1}^{J}\sum_{k=1}^{H_{ij}} \text{E}_{U|X,\theta_t}[u_iu_i^T],
\end{equation*}
\begin{equation*}
\Sigma_v =  \frac{1}{\sum_{i=1}^I\sum_{j=1}^{J}\sum_{k=1}^{H_{ij}}1}\sum_{i=1}^I\sum_{j=1}^{J}\sum_{k=1}^{H_{ij}} \text{E}_{V|X,\theta_t}[v_jv_j^T],
\end{equation*}
\begin{eqnarray*}
 \Sigma_\epsilon &=&\frac{1}{\sum_{i=1}^I\sum_{j=1}^{J}\sum_{k=1}^{H_{ij}}1} \sum_{i=1}^I\sum_{j=1}^{J}\sum_{k=1}^{H_{ij}} \{(x_{ij}-\mu)(x_{ijk}-\mu)^T\\&-&2(x_{ijk}-\mu)[\text{E}_{U|X,\theta_t}[u_i]^T+\text{E}_{V|X,\theta_t}[v_i]^T] \\
 &+& \left(\text{E}_{U|X,\theta_t}[u_iu_i^T]
+2\text{E}_{U,V|X,\theta_t}[u_iv_j^T]+\text{E}_{V|X,\theta_t}[v_jv_j^T]\right)
 \}, \nonumber
\end{eqnarray*}
and
\begin{equation*}
\mu = \frac{\sum_{i=1}^I\sum_{j=1}^{J}\sum_{k=1}^{H_{ij}}  x_{ijk}}{\sum_{i=1}^I\sum_{j=1}^{J}\sum_{k=1}^{H_{ij}}  1}.
\end{equation*}
The expectation terms $\text{E}_{U|X,\theta_t}[u_i]$, $\text{E}_{V|X,\theta_t}[v_j]$, $\text{E}_{U|X,\theta_t}[u_iu_i^T]$, $\text{E}_{V|X,\theta_t}[v_jv_j^T]$, and  $\text{E}_{U,V|X,\theta_t}[u_iv_j^T]$ can be extracted from Equations~\eqref{eq:estep1},~\eqref{eq:estep2} and~\eqref{eq:estep3}.

\subsection{Likelihood Ratio Scores}

We treat the verification as a kind of hypothesis testing problem with the null hypothesis $\mathcal{H}_0$ where two j-vectors have the same speaker and phrase variables $u_i$ and $v_{j}$ and the alternative hypothesis $\mathcal{H}_1$ where they do not (there are three cases: different underlying $u_i$ variable with same $v_{j}$ variable in model $\mathcal{M}_1$, same $u_i$ variable with different $v_{j}$ variables in model $\mathcal{M}_2$, or different underlying $u_i$ variables with different $v_{j}$ variables in model $\mathcal{M}_3$).
Given a test j-vector $x_t$ and an enrolled j-vector $x_s$, and let the priori probability of the models $\mathcal{M}_1$, $\mathcal{M}_2$, $\mathcal{M}_3$ as $p_1=P(\mathcal{M}_1|\mathcal{H}_1)$, $p_2=P(\mathcal{M}_2|\mathcal{H}_1)$, $p_3=P(\mathcal{M}_3|\mathcal{H}_1)$, then the likelihood ratio score is
\begin{align*}
 &l(x_t,x_s)= \frac{P(x_t,x_s|\mathcal{H}_0)}{P(x_t,x_s|\mathcal{H}_1)}\\=&\frac{\int\int p(x_t,x_s|u_1,v_1,\theta)p(u_1)p(v_1)du_1dv_1}{\textbf{X}} \\
=&\frac{\mathcal{N}(\left[\begin{matrix}
    x_t\\ x_s
    \end{matrix}\right]|\left[\begin{matrix}
    \mu\\ \mu
    \end{matrix}\right],\left[\begin{matrix}
    \Sigma_u + \Sigma_v  + \Sigma_\epsilon & \Sigma_u + \Sigma_v   \\ \Sigma_u + \Sigma_v   & \Sigma_u + \Sigma_v  +\Sigma_\epsilon
    \end{matrix}\right])}{\textbf{X}},
\end{align*}
where
\begin{align*}
 &\textbf{X}=P(x_t,x_s|\mathcal{H}_1)= P(x_t,x_s|\mathcal{M}_1)P(\mathcal{M}_1|\mathcal{H}_1)
\\+&P(x_t,x_s|\mathcal{M}_2)P(\mathcal{M}_2|\mathcal{H}_1)+P(x_t,x_s|\mathcal{M}_3)P(\mathcal{M}_3|\mathcal{H}_1)\\=&p_1\int\int\int  p(x_t,x_s,u_1,u_2,v_1|\theta)du_1du_2dv_1\\+&p_2\int\int\int  p(x_t,x_s,u_1,v_1,v_2|\theta)du_1dv_1dv_2\\ +&p_3\int\int  p(x_t,u_1,v_1|\theta)du_1dv_1\int\int  p(x_s,u_2,v_2|\theta)du_2dv_2 \\
=&p_1\mathcal{N}(\left[\begin{matrix}
    x_t\\ x_s
    \end{matrix}\right]|\left[\begin{matrix}
    \mu\\ \mu
    \end{matrix}\right],\left[\begin{matrix}
  \Sigma_u + \Sigma_v +\Sigma_\epsilon & \Sigma_v\\  \Sigma_v & \Sigma_u + \Sigma_v +\Sigma_\epsilon
    \end{matrix}\right])\\+&p_2\mathcal{N}(\left[\begin{matrix}
    x_t\\ x_s
    \end{matrix}\right]|\left[\begin{matrix}
    \mu\\ \mu
    \end{matrix}\right],\left[\begin{matrix}
    \Sigma_u + \Sigma_v +\Sigma_\epsilon & \Sigma_u   \\ \Sigma_u   &\Sigma_u + \Sigma_v +\Sigma_\epsilon
    \end{matrix}\right])\\ +& p_3\mathcal{N}(x_t|\mu, \Sigma_u + \Sigma_v +\Sigma_\epsilon)\mathcal{N}(x_s|\mu, \Sigma_u + \Sigma_v +\Sigma_\epsilon).
\end{align*}

Notice that like standard joint Bayesian model~\cite{chen2017efficient}, we do not calculate a point estimate of hidden variable. Instead we compute the probability that the two multi-label vectors had the same hidden variables, regardless of what this actual latent variable was.

\section{Experiments}

In this section, we describe the experimental setup and results for the proposed method on the public RSR2015 English corpus~\cite{larcher2014text} and our internal Huiting202 Chinese Mandarin database collected by the Huiting Techonogly\footnotemark[2].

\footnotetext[2]{http://huitingtech.com/}

\subsection{Experimental setup}

RSR2015 corpus~\cite{larcher2014text} was released by I2R, is used to evaluate the performance of different speaker verification systems. In this work, we follow the setup of~\cite{liu2015deep}, the part I of RSR2015 is used for the testing of DoJoBa. The background and development data of RSR2015 part I are merged as new background data to train the j-vector extractor.

Our internal gender balanced Huiting202 database is designed for local applications. It contains 202 speakers reading 20 different phrases, 20 sessions each phrase. All speech files are of 16kHz. 132 randomly selected speakers are used for training the background multi-task learned DNN, and the remaining 70 speakers were used for enrollment and evaluation.

In this work, 39-dimensional Mel-frequency cepstral coefficients (MFCC, 13 static including the
log energy + 13 $\Delta$ + 13 $\Delta \Delta$) are extracted and normalized using utterance-level
mean and variance normalization. The input is stacked normalized MFCCs from 11 frames (5 frames from
each side of the current frame).
The DNN has 6 hidden layers (with sigmoid activation
function) of 2048 nodes each. During the background model development stage, the DNN was trained by the strategy of pre-training with Restricted Boltzmann Machine (RBM) and fine tuning with SGD using cross-entropy criterion.
Once the DNN is trained, the j-vector can be extracted during the enrollment and evaluation stages.

\subsection{Results and discussion}

Four systems are evaluated and compared across above conditions:
\begin{itemize}
\item
\textbf{j-vector}: the standard j-vector system with cosine similarity.
\item
\textbf{joint Bayesian}: the j-vector system with classic joint Bayesian in~\cite{chen2017efficient}.
\item
\textbf{jPLDA}: joint PLDA system described in~\cite{Shi2017Multi} with j-vector.
\item
\textbf{DoJoBa}: double joint Bayesian system described in Section~\ref{sec:dojoba} with j-vector.
\end{itemize}

When evaluation a speaker is enrolled with 3 utterances of the same phrase. The task concerns on both the phrase content and speaker identity.
Nontarget trials are of three types: the impostor pronouncing wrong lexical content (impostor wrong, IW); a target speaker pronouncing wrong lexical content (target wrong, TW); the imposter pronouncing correct lexical content (impostor correct, IC).

The joint Bayesian, jPLDA, and DoJoBa models are trained using the j-vectors. The class defined in both models is the multi-task label of both the speaker and phrase. For each test session the j-vector is extracted using the same process and then the log likelihood from joint Bayesian, jPLDA, and DoJoBa are used to distinguish among different models. The number of principle components is set to 100 and then the joint Bayesian model is estimated  with 10 iterations; the speaker and the phrase subspace dimensions of jPLDA and DoJoBa are both set to 100 regarding of fair comparisons and the jPLDA and DoJoBa model are also trained with 10 iterations.

Table~\ref{tab:rsr2015} and~\ref{tab:huiting202} compare the performances of all above-mentioned systems in terms of equal error rate (EER) for the three types of nontarget trials. Obviously DoJoBa is superior to the standard joint Bayesian and jPLDA, regardless of the test database.
Since DoJoBa system can explore both the identity and the lexical information from the j-vector, it constantly performs better than standard joint Bayesian systems.

\begin{table}[th]
\caption[rsr2015]{Performance of different systems on the evaluation set of RSR2015 part I in terms of equal error rate (EER \%).}\label{tab:rsr2015}
\centering
\begin{tabular}{|c|c|c|c|c|}
\hline
EER(\%) & j-vector &  joint Bayesian & jPLDA & DoJoBa \\
\hline
IW & 0.95& 0.02 &   0.02 & 0.02\\
\hline
TW & 3.14 & 0.03 & 0.06 & 0.02\\
\hline
IC & 7.86 & 3.61 &  3.12 & 2.97\\
\hline
Total & 1.45 & 0.46 & 0.40 & 0.37\\
\hline
\end{tabular}
\end{table}

\begin{table}[th]
\caption[huiting202]{Performance of different systems on the evaluation set of Huiting202 in terms of equal error rate (EER \%).}\label{tab:huiting202}
\centering
\begin{tabular}{|c|c|c|c|c|}
\hline
EER(\%) & j-vector &  joint Bayesian & jPLDA & DoJoBa \\
\hline
IW & 0.86& 0.10 &  0.13 & 0.08\\
\hline
TW & 6.71 & 0.04 & 0.07 & 0.04\\
\hline
 IC &  4.57 & 2.52 &   2.37& 2.13\\
\hline
Total & 1.37 & 0.45 &  0.36 & 0.31 \\
\hline
\end{tabular}
\end{table}

\section{Conclusions}

In this paper we have proposed a double joint Bayesian (DoJoBa) analysis for j-vector verification. DoJoBa is related to joint Bayesian model, and can be thought of as joint Bayesian with multiple probability distributions attached to the features. The most important advantages of DoJoBa, compared to joint Bayesian, is that multiple information can be explicitly modeled and explored from the samples to improve the verification performance; comparing to jPLDA, DoJoBa can determine the latent dimension without tuning. Reported results showed that DoJoBa provided significant reduction in error rates over conventional systems in term of EER.

\bibliographystyle{IEEEbib}
\bibliography{dojoba}

\end{document}